\documentclass[9pt,twocolumn,twoside]{pnas-new}

\templatetype{pnasresearcharticle} 

\usepackage{bm}
\usepackage{pdfpages}
\usepackage{placeins}
\let\oldequation\equation
\let\oldendequation\endequation

\renewenvironment{equation}
  {\linenomathNonumbers\oldequation}
  {\oldendequation\endlinenomath}

\let\oldalign\align
\let\oldendalign\endalign

\renewenvironment{align}
  {\linenomathNonumbers\oldalign}
  {\oldendalign\endlinenomath}
  
\def\be{\begin{equation}}
\def\ee{\end{equation}}

\DeclareMathOperator{\tr}{Tr}
\DeclareMathOperator{\kl}{KL}
\DeclareMathOperator*{\argmax}{argmax}
\def\bmh{\bm{h}}
\def\bmj{\bm{J}}
\def\bms{\bm{s}}
\def\trinvi{\tr \mathcal{I}^{- 1}}
\newcommand{\bangle}[1]{\left\langle #1 \right\rangle}
\newcommand{\pfrac}[2]{\frac{\partial #1}{\partial #2}}
\newcommand{\pfracsec}[3]{\frac{\partial^2 #1}{\partial #2 \partial #3}}

\title{Active Learning of Spin Network Models}

\author[a,1]{Jialong Jiang}
\author[b]{David A. Sivak} 
\author[a,1]{Matt Thomson}

\affil[a]{Division of Biology and Biological Engineering, California Institute of Technology, Pasadena, CA 91125}

\affil[b]{Department of Physics, Simon Fraser University, Burnaby, BC V5A 1S6, Canada}

\leadauthor{Jiang} 

\significancestatement{Learning direct interaction between nodes in complex networks from observed correlation is always difficult. In the natural sciences, applying interventions is crucial to uncover such interactions. The development of new technologies now enables the performance of high-throughput perturbation experiments to probe network structure. However, the analysis and design of such experiments still remains mostly empirical. We develop an inference framework that allows automated and active learning of statistical models through iterative rounds of observation and perturbation based on Fisher information and sequential Bayesian inference. 
We show that the perturbations break strong correlations between variables, thus reducing the sampling complexity of the inference by orders of magnitude. 
The framework also raises new statistical problems for further investigation. 
}

\authorcontributions{M.T. and D.S. conceived of the idea. J.J. and M.T. developed the theory and performed the computations. J.J. developed the Bayesian inference framework. J.J. and M.T. wrote the manuscript. }
\authordeclaration{The authors declare no conflict of interest.}
\correspondingauthor{\textsuperscript{1}To whom correspondence should be addressed. E-mail: mthomson@caltech.edu or jiangjl@caltech.edu}


\begin{abstract}
The inverse statistical problem of finding direct interactions in complex networks is difficult. In the natural sciences, well-controlled perturbation experiments are widely used to probe the structure of complex networks. However, our understanding of how and why perturbations aid inference remains heuristic, and we lack automated procedures that determine network structure by combining inference and perturbation. 
Therefore, we propose a general mathematical framework to study inference with iteratively applied perturbations. Using the formulation of information geometry, our framework quantifies the difficulty of inference and the information gain from perturbations through the curvature of the underlying parameter manifold, measured by Fisher information. We apply the framework to the inference of spin network models and find that designed perturbations can reduce the sampling complexity by $10^6$-fold across a variety of network architectures. Physically, our framework reveals that perturbations boost inference by causing a network to explore previously inaccessible states. Optimal perturbations break spin-spin correlations within a network, increasing the information available for inference and thus reducing sampling complexity by orders of magnitude. Our active learning framework could be powerful in the analysis of complex networks as well as in the rational design of experiments. 
\end{abstract}

\dates{This manuscript was compiled on \today}

\begin{document}

\maketitle
\thispagestyle{firststyle}
\ifthenelse{\boolean{shortarticle}}{\ifthenelse{\boolean{singlecolumn}}{\abscontentformatted}{\abscontent}}{}

\dropcap{A}
significant property of complex systems is the convoluted interaction between different parts. 
Describing the structure of interactions in a network is critical to understanding and predicting its behavior. Numerous models have been developed to characterize complex networks, and many different methods are used to infer network interactions from the data generated by a network, for example, methods based on variable correlation, mutual information between variables, likelihood, and temporal dynamic relationships~\citep{le2015quantitative}.

However, difficulties are always confronted while solving the problem of deducing direct interaction from correlation, as many alternative causal relations can all 
explain the same observed correlations. 
Many disciplines in scientific research, especially biology, rely on perturbations to tackle the inference problem. 
For example, gene functions are studied by their mutants, and signaling pathways are decoded from carefully designed knock-in/out experiments. 
Recent developments in molecular biology provide high-throughput technology to perform perturbation experiments, such as CRISPR/Cas9 in gene editing~\citep{hsu2014development, dixit2016perturb} and optogenetics in neuron activity control~\citep{deisseroth2011optogenetics}. 
With these methods, it is natural to ask, in general, how to design perturbation experiments to make the most accurate and efficient inference, and how perturbation can be helpful in deciphering the networks.

There have been studies of optimal design~\citep{paninski2005asymptotic, hyttinen2013experiment} and analysis~\citep{molinelli2013perturbation} of perturbation experiments, and efforts to connect perturbation and inference in an iterative process~\citep{ideker1999discovery}.
Active learning of Bayesian networks on directed acyclic graphs has been studied from many facets in causal inference.
Interventions are modeled as pinning down node values to distinguish between Markov-equivalent models~\cite{murphy01activelearning, tong2001active, he2008active, cho2016reconstructing}. 
Information-theoretic measures have been developed in the field of optimal experimental design~\citep{pukelsheim2006optimal, atkinson2007optimum, jeong2018experimental, apgar2010sloppy, transtrum2012optimal}, to select optimal observation times, sets of variables for measurement, or experimental conditions. 
Nonetheless, previous works address rather specific classes of models, or only provide an empirical numerical procedure without quantitative insight into how well active learning works and why it works. 
A general mathematical framework to understand how perturbations facilitate inference of physical models is still absent.
Specifically, we have yet to understand to what extent perturbations can reduce the sampling complexity of the original inference problem or why, physically, performing a perturbation is helpful. 

To demonstrate our framework, we constrain ourselves to a specific and canonical class of network models, the spin networks as probabilistic graphical models. 
Nonetheless, the framework developed in this paper can be generalized to other probabilistic models without difficulty~(\textit{SI Appendix}, Text 1A). 

Spin networks, or spin glasses, represent a broad class of networks with interesting physical behaviors such as multi-stability.
Spin glasses have provided a canonical mathematical framework for understanding and analyzing properties of complex interacting systems across many disciplines ranging across computational biology~\citep{marks2011protein, lezon2006using}, neuroscience~\citep{cocco2009neuronal}, and data science~\citep{nguyen2017inverse, hinton2006fast}. 
The inference problem involved in parametrizing a spin network model from data is known in physics as the inverse Ising problem or spin glass inverse problem.
To solve the inverse Ising problem is formally hard, both in sampling complexity and computational cost~\citep{nguyen2017inverse, santhanam2012information, montanari2009graphical}. 
Formally, detailed analysis of sampling complexity shows that the number of samples needed to distinguish different structures grows polynomially with the number of edges, but exponentially with the $ \ell_\infty $ norm of interaction matrix $ \bmj $, which represents the coupling strength between nodes~\citep{santhanam2012information}.
However, these properties have not been investigated in the setting of active learning through designed perturbations. 

In this paper, we propose a framework to perform parametric estimation of a spin network with the ability to perturb the system so that we can iteratively update our knowledge through different perturbation experiments. 
In the context of the inverse Ising problem, we learn the coupling matrix $ \bmj $ while controlling the field term $ \bmh $. 
We demonstrate procedures for designing experiments and a learning process to achieve significant improvement in inference accuracy on medium-sized networks with strong couplings. 
We show that optimal perturbations disrupt strong spin-spin correlations within a network, altering the eigenspectrum of the Fisher information matrix, leading to order-of-magnitude reductions in sampling complexity. 
Across a broad range of network topologies, perturbations reduce the sampling complexity of inference by orders of magnitude when compared with `passive' inference.
These new insights into the performance of active learning for spin network models lead to further questions in statistics and potential applications to complex networks across the natural sciences. 

\section*{Theoretical Framework}

\subsection*{Iterative inference with perturbation}

A spin network is a probabilistic graphical model with each node taking 
a
value in $ \{1, - 1\} $.
For a $ p $-node network, the probability distribution over $ 2^p $ configurations is the induced Boltzmann distribution from an Ising-type interaction energy. 
Then the probability of a configuration $ \bms $ given an interaction matrix $ \bmj $ and a local field $ \bmh $ is
\begin{align}
&P(\bms | \bmj, \bmh) = \frac{\exp[- E(\bms)]}{\mathcal{Z}} \, ,\nonumber \\
&E(\bms) = - \sum_{i < j} J_{ij} s_i s_j - \sum_i h_i s_i \, , 
\end{align} 
where $ \mathcal{Z} = \sum_{\{\bms\}} \exp[- E(\bms)] $ is the partition function. 
For simplicity, the inverse temperature factor $ \beta $ is absorbed into the parameters for interactions and field strengths. 
In general, the learning or inference of the network model consists of finding the best $ \bmj $ and $ \bmh $ to describe the observed data, which can be solved by maximum-likelihood estimation (MLE), pseudo-likelihood~\citep{aurell2012inverse} or other approximate optimization methods~\citep{vuffray2016interaction}.

In our framework, we consider the scenario where $ \bmj $ is unknown but $ \bmh $ can be manipulated to facilitate inference of $ \bmj $. 
In a spin network, the local field $ \bmh $ describes a tendency of activation for every node. Therefore, we allow a local magnetic field to bias each spin during inference. 
The magnetic field models a common type of perturbations that activates/deactivates each node individually, such as knockdown of genes or induced activation of neurons. 
For simplicity, we assume that we have full control of the field, namely the system does not have an unknown intrinsic field. 
The case with an intrinsic field can be dealt with similarly using our framework. 

In practice, a single perturbation can be insufficient, and inference is performed across a combination of different perturbations. 
Therefore, samples are taken with different fields $ \bmh^{(1)}, \bmh^{(2)}, \ldots, \bmh^{(n)}$, and the information is integrated through the following sequential Bayesian estimation. 
Specifically, we seek the maximum-likelihood estimator $ \tilde{\bmj} = \argmax_{\bmj'} \, \log P(\{\bms\} | \bmj', \bmh) $ for a spin network and solve the optimization by gradient ascent. 
To combine information across multiple rounds of perturbation, 
MLE can be alternatively viewed as maximizing a Bayesian posterior.
Hence, the inference with different fields can be performed sequentially by
updating the posterior with the current round of data while using the posterior of all previous rounds as prior. 
The gradient of the log posterior can be computed by the recursive formula 
\be  
\pfrac{\log P_n}{J_{ij}} = \bangle{s_i s_j}_n^{\text{o}} - \bangle{s_i s_j}_n + \pfrac{\log P_{n - 1}}{J_{ij}} \, ,
\ee 
where $ P_n $ is the posterior distribution given all the samples under the first $ n $ fields (\textit{SI Appendix}, Text 1B).
$ \bangle{s_i s_j}^{\text{o}} $ is the average over the observed samples $ \{\bms\} $, and $ \bangle{s_i s_j} $ is the average over the distribution generated by the current parameters. 
Even with this closed form, exact evaluation of the gradient involves exponentially many terms and is generally approximated by samples taken from Markov-chain Monte Carlo sampling. 
The computational cost of the Bayesian gradient only depends linearly on the number of learning rounds. 
The log likelihood is additive, so the final landscape defined by the log posterior is the sum of the landscapes for each individual perturbation. 

\subsection*{Selecting perturbations by Fisher information}

We use measures from information geometry to quantify the difficulty of inference and design perturbations to facilitate the inference. 
Information geometry defines a geometric structure to characterize the change in a probability distribution with changes in underlying parameters. 
For a parametric family of distributions $ P(\bm{x} | \bm{\theta}) $, the difference between any two distributions measured by Kullback-Leibler divergence can be expanded in the differential change of parameters $ \delta \bm{\theta}$
\begin{align}
\kl(P(\bm{x} | \bm{\theta}), P(\bm{x} | \bm{\theta} + \delta \bm{\theta})) &= \frac{1}{2} \delta \bm{\theta}^T \mathcal{I} \delta \bm{\theta} + \mathcal{O}(\delta \bm{\theta}^3) \nonumber \\ 
\mathcal{I} = - \bangle{\pfracsec{\log P(\bm{x} | \bm{\theta})}{\theta_i}{\theta_j}} &= \bangle{\pfrac{\log P}{\theta_i} \pfrac{\log P}{\theta_j}} \, .
\end{align}
For inverse Ising inference, the Fisher information matrix (FI) can be derived from properties of the exponential family of distributions
\begin{align}
\label{eq:fi}
I_{\{ij\}, \{kl\}} &= \bangle{s_i s_j s_k s_l} - \bangle{s_i s_j}\bangle{s_k s_l} \,,
\end{align}
where $ \{ij\} $ corresponds to interaction term $ J_{ij} $.
The FI $ \mathcal{I} $ is a Riemannian metric that describes how the parametric density manifold curves:
small FI corresponds to a small change in the probability distribution given a change in parameter values, making the inference difficult. 
In our framework, FI is computed with respect to $ \bmj $ given fixed $ \bmh $, so that FI is a function of the applied field. 

The Cram\'er–Rao bound relates the Fisher information matrix of a given system to the minimum inference error of estimators. 
Specifically, the covariance $ C \succeq \mathcal{I}^{- 1} $ for any unbiased estimator, in the sense of $ C - \mathcal{I}^{- 1} $ being positive semidefinite. 
Moreover, FI is the expectation of the Hessian matrix of the log-likelihood function, the condition number of which quantifies the difficulty of numerical optimization of the MLE~\citep{watanabe2009algebraic}. 

Furthermore, the Cram\'er–Rao bound connects FI with the sampling complexity of a given inference problem. 
The FI of $ N $ independent samples is $ N \mathcal{I} $, hence $ \Omega (\epsilon^{- 1} \lambda^{- 1}) $ samples are needed to achieve error $ \epsilon $ in expectation on the projection of the parameters onto the eigenvector of FI with eigenvalue $ \lambda $. 
Furthermore, $ \Omega (\epsilon^{- 1} \tr \mathcal{I}^{- 1}) $ samples 
are required for $ \ell_2 $ error smaller then $ \epsilon ^ 2 $. 
Therefore, we set $ \trinvi $ as the objective function while optimizing $ \bmh $.
Assuming that the number of samples is the same for each round, and that perturbations are chosen sequentially, the optimal choice of $ \bmh $ in the $ n $-th round is 
\begin{align}
\label{eq:hn}
\min_{\bmh_n} \quad &\tr \ \mathcal{I}^{- 1}_n \nonumber \\
\mathrm{s.t.} \quad &\mathcal{I}_i = \mathcal{I}_{i - 1} + \mathcal{I}(\bmj, \bmh_i) \quad i = 1, \ldots, n \\
&\mathcal{I}_0 = 0 \nonumber \,,
\end{align}
where $ \mathcal{I}_n $ is defined recursively. 

However, when using the method to uncover the structure of networks in applications, $ \mathcal{I}(\bmj, \bmh_i) $ cannot be evaluated directly as $ \bmj $ is unknown. So we need to approximate $ \mathcal{I}(\bmj, \bmh_n) $ with our current estimate $ \tilde{\bmj} $.
For $ i = 1, \ldots, n - 1 $, we already acquired samples from the real system, thus $ \mathcal{I}_i $ can be approximated using the empirical distribution instead of the true distribution in Eq.~\ref{eq:fi}. 
The procedure runs in an iterative way between computation and experiments: new perturbations are designed based on previous samples and the resulting estimate $ \tilde{\bmj} $. 
Each time new samples are taken from the system, we construct a new estimate of $ \tilde{\bmj} $, and solve the optimization Eq.~\ref{eq:hn} to define the most informative perturbation to execute in the next experiment. 
This framework can also be expanded to perform multiple new perturbations $ \bmh_n $ each round.

\subsection*{Examples: two-spin network and Ising chain}

Analytical examples provide insights on the sampling complexity of the active inference and how perturbations help to improve the FI eigenvalue spectrum.
Results in the literature for passive inference show that sampling complexity grows exponentially with the magnitude of entries in $ \bmj $~\cite{santhanam2012information}. 
The difficulty of inferring a spin network largely comes from strong coupling between nodes, making different network structures indistinguishable. 
External fields can break such correlations, decrease $ \trinvi $ even to its lower bound, thus reducing the sampling complexity. 
The lower bound of $ \trinvi $ is realized when $ \mathcal{I} $ is the identity matrix (\textit{SI Appendix}, Text 2A).
In the following special cases, suitable perturbations maintain nearly-constant
sampling complexity with increasing $ \bmj $.

\begin{figure*}[t!]
\centering
\includegraphics[width=17.8cm]{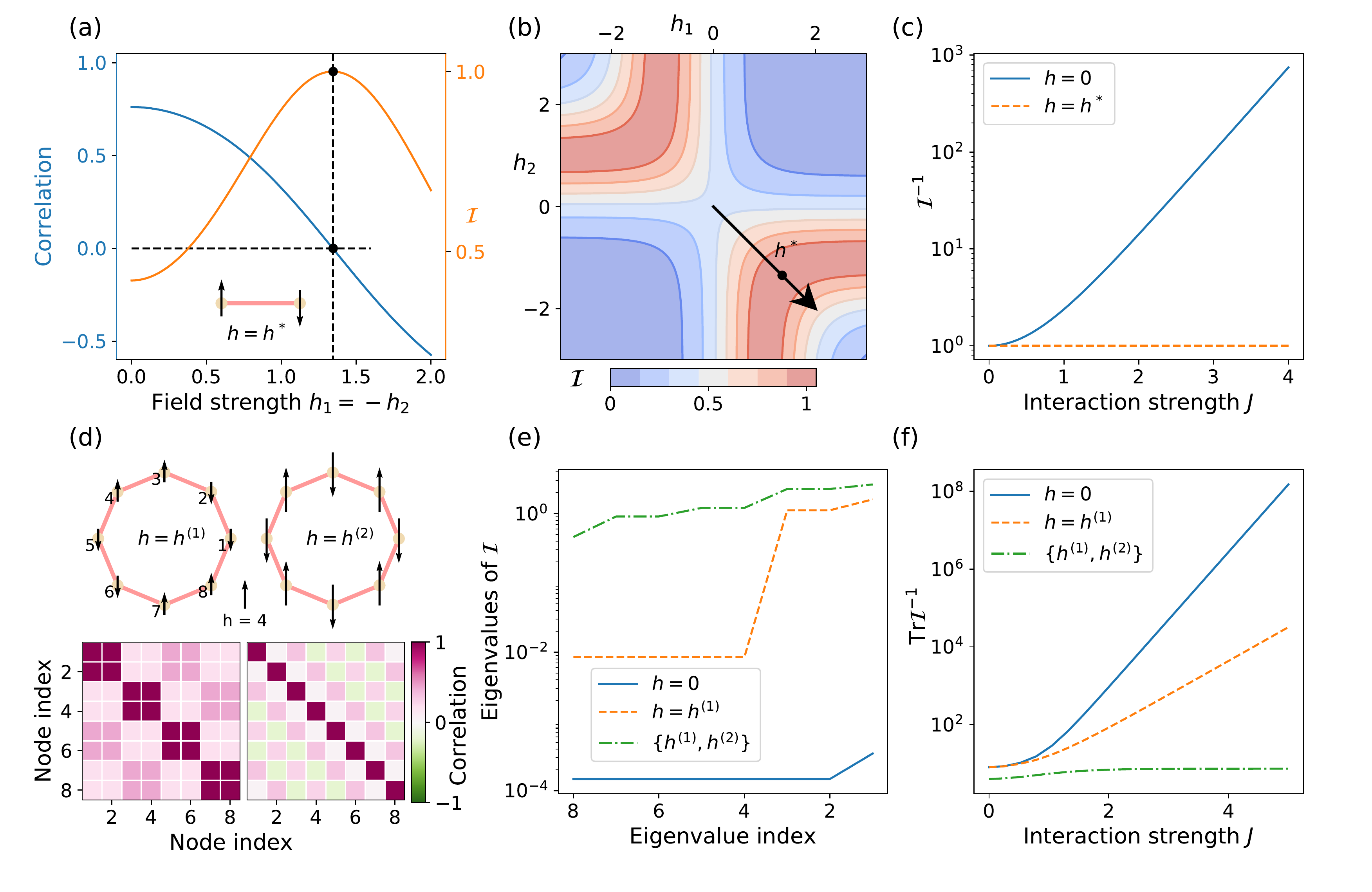}
\caption{\textbf{Perturbations and scaling of two-node inference and Ising-chain inference.}
The interaction strength $ J = 1 $ in (a),(b). 
(a) The inset shows the two-spin network with the direction of a optimal perturbation $ h_1 = - h_2 $.
Correlation and Fisher information $ \mathcal{I} $ change with different field strength along this direction. 
$ \mathcal{I} $ achieves its maximum $ 1 $ when correlation is $ 0 $. 
(b) Landscape of the Fisher information $ \mathcal{I} $ with different perturbation fields $ h_1, h_2 $.
The arrow shows the perturbation direction used in (a). 
(c) Scaling of $ \tr \mathcal{I}^{- 1} $ in two-node networks with interaction strength $ J $, without perturbation and with field $ \bmh^{*} $.
\hspace{2em} 
The interaction strength $ J = 3 $ in (d),(e).
(d) Illustration of the 8-node ferromagnetic Ising chain, and its correlation matrices under two different perturbations $ \bmh^{(1)} $ and $ \bmh^{(2)} $. 
(e) Eigenvalues of FI 
for the Ising chain with no perturbation, with field $ \bmh^{(1)} $, and 
with the sequential application of $ \{ \bmh^{(1)}, \bmh^{(2)} \} $. 
(f) Scaling of $ \tr \mathcal{I}^{- 1} $ in the Ising chain with interaction strength $ J $, for no perturbation, with field $ \bmh^{(1)} $, and with sequential application of $ \{ \bmh^{(1)}, \bmh^{(2)} \} $.
}
\label{fig:ana_sol}
\end{figure*}

For the simplest case of a two-spin network, FI is a scalar.
So the optimum corresponds to the maximum of $ \mathcal{I} $, where $ \bangle{s_1 s_2} = 0 $, achieved by (\textit{SI Appendix}, Text 2B)
\be 
h_2 = \frac{1}{2} \log \frac{1 - \exp(2 J + 2 h_1)}{\exp 2J - \exp 2 h_1} \,.
\ee
The optimal $ \mathcal{I} = 1 $ can be achieved by an infinite number of $ (h_1, h_2) $, and one special approximate solution is $ h_1 = - h_2 = J + \log \sqrt{2} $ for large positive $ J $, as shown in Fig.~\ref{fig:ana_sol}~(a) where $ J = 1 $. 
The landscape of $ \mathcal{I} $ as a function of $ h_1, h_2 $ is shown in Fig.~\ref{fig:ana_sol}~(b).
Note that this landscape is nonconvex, and there is no unique maximum.
The point without field $ (h_1 = 0, h_2 = 0) $ is a saddle point, with two principal-axis directions $ (1, 1) $ and $ (1, - 1) $.
Without the field ($ \bmh = 0 $) , $ \mathcal{I}^{ - 1} = 1 / (1 - \tanh ^ 2 J) = \cosh ^ 2 J $, which means that sampling complexity increases exponentially with $ J $.
In contrast, an optimal field $ \bmh^* $ produces $ \trinvi = 1 $, so the sampling complexity is reduced exponentially, as shown in Fig.~\ref{fig:ana_sol}~(c).

Intuitively, the difficulty of inference is caused by the high probability of 
the two states $ [1, 1], [- 1, - 1] $ and the corresponding small probability of the other two states, producing correlation $ 1 $ of the two spins. 
Fields in the direction of $ (1, - 1) $ make one of the previous low-probability states more accessible, thus increasing FI. 
On the other hand, 
fields in the direction of $ (1, 1) $ further concentrate the distribution, decreasing the FI. 
Thus $(h_1 = 0, h_2 = 0)$ forms a saddle point. 

Another canonical model is the finite ferromagnetic Ising chain with periodic boundary conditions. 
For an analytical solution, we restrict ourselves to the case of knowing the chain structure and inferring the magnitude of individual interaction strengths. 
The energy function is 
\be 
E = - \sum_{i = 1}^p J_i s_i s_{i + 1} - \sum_{i = 1}^p h_i s_i \, ,
\ee
where the convention of $ s_{p + 1} \equiv s_1 $ is used. 
The FI can be solved approximately when $ J_i $ are all equal to $ J > 0 $, and $ h_i = 0 $ (\textit{SI Appendix}, Text 2C):
\begin{align}
\mathcal{I}_{\{i, i + 1\}\{j, j + 1\}} =
\begin{cases}
4 (p - 1) \exp(- 4 J) & i = j \\
4 \exp(- 4 J) & i \neq j
\end{cases}
\end{align}
The FI is a circular matrix so its eigenvectors have the form $ (1, \omega_j, \ldots, \omega_j^{p - 1}) $, where $ \omega_j = \exp(j \ 2 \pi i / p) $. 
There is one large eigenvalue $ \lambda_1 = 8 (p - 1) \exp(- 4 J) $ and $ (p - 1) $ degenerate small eigenvalues $ \lambda_2 = 4 (p - 2) \exp(- 4 J) $, as shown in Fig.~\ref{fig:ana_sol}~(b).

We choose two perturbations $ h^{(1)}_j = h_0^{(1)} - (- 1)^{\lfloor j / 2 \rfloor} $ and $ h^{(2)}_j = h_0^{(2)} (- 1)^{j} $, where $ \lfloor j / 2 \rfloor $ is the largest integer less than $ j / 2 $.
$ h^{(1)}_0, h^{(2)}_0 $ are obtained by numerical minimizing $ \trinvi $. 
Fig.~\ref{fig:ana_sol}~(d)
illustrates the perturbations for the uniform $ J = 3 $ chain.
Without field, the two states corresponding to all $ 1 $s or all $ - 1 $s dominate the distribution, producing a correlation matrix with all entries near $ 1 $.
The two fields break the correlation in different ways, as shown in Fig.~\ref{fig:ana_sol}~(d). 

The eigenvalues of FI under
$ h^{(1)} $ and the combination $ \{h^{(1)}, h^{(2)} \} $ are shown in Fig.~\ref{fig:ana_sol}~(e). 
With $ h^{(1)} $ alone, all eigenvalues are increased but there are still relatively small ones. 
The combination $ \{h^{(1)}, h^{(2)} \} $ raises all eigenvalues to 
around $ 1 $.
Even more striking is the effect on the scaling of $ \trinvi $ with interaction intensity $ J $: $ \trinvi $ grows exponentially with $ J $ under no
perturbation or only $ \bmh^{(1)} $, while it remains near-constant under the combination $ \{h^{(1)}, h^{(2)} \} $, as shown in Fig.~\ref{fig:ana_sol}~(f).
Thus the combination of perturbations reduces the sampling complexity from exponential to almost constant. 

This example shows that a single perturbation sometimes is not sufficient to produce large eigenvalues for all eigenvectors (hence easy inference). 
Effects of perturbation strongly depend on network structure, as illustrated in a complete analysis of three-node networks~(\textit{SI Appendix}, Text 3, Fig. S1$-$3).
However, as FI is additive for independent samples, we can combine the information from many samples with different choices of fields. 
In the geometric viewpoint, eigenvectors with small eigenvalues in the FI represent singular directions like flat valleys 
with small second-order derivative near the maximum and, therefore, low local curvature in the likelihood landscape. 
Combining samples taken from different conditions is equivalent to adding these landscapes together; 
as the singular dimensions differ under different perturbations, combining the landscapes can make the overall landscape strongly curved in all directions. 

\section*{Numerical Results}

\subsection*{Inference of a modular spin network}

In larger networks, analytical solutions are unavailable but we can numerically study the impact of perturbations on active learning through the Eq.~\ref{eq:hn}. 
We demonstrate good perturbations exist that can decrease $ \trinvi $ by orders of magnitude, by calculating the optimal perturbations with the correct model of the underlying network. 

\begin{figure*}[t!]
\centering
\includegraphics[width=17.8cm]{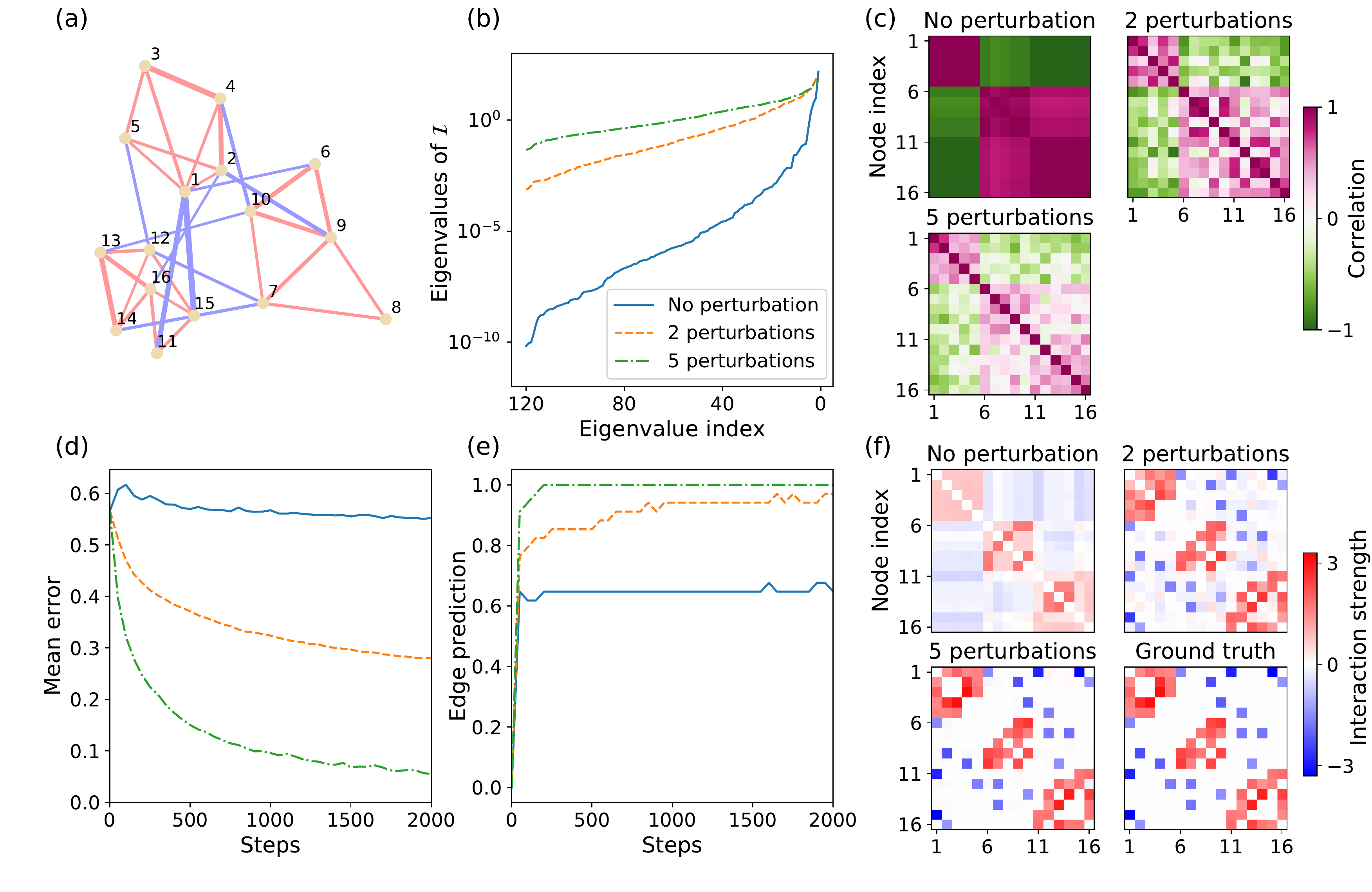}
\caption{\textbf{Inference of a modular spin networks} 
(a) 
Network structure to be inferred. 
Red edges represent $ J_{ij} > 0 $ and blue edges represent $ J_{ij} < 0 $. 
Each edge width is proportional to $ |J_{ij}| $. 
(b) Eigenvalue spectrum of Fisher information 
matrix for the inference problem, without perturbation or for different
numbers of perturbations. 
(d),(e) share the same legend as (b). 
(c) Correlation matrices for different numbers of perturbations. 
(d) Mean estimation error of $ J_{ij} $ as a function of training steps. 
Each experiment has the same total number of samples.
(e) Edge prediction precision 
(fraction of $K$ true edges $\bmj$ present in strongest $K$ edges from inference) 
as a function of training steps. 
(f) The estimation $ \tilde{\bmj} $ with different number of perturbations and true interaction matrix $ \bmj $.
}
\label{fig:oracle}
\end{figure*}

In many real systems, networks are composed of several communities or modules~\citep{newman2006modularity}.
One common form is a network that has activation inside each module and repression between different modules. 
For such systems, samples obtained from the unperturbed network 
are always inadequate to infer the sign and exact strength of these interactions. 
We performed our method on a 16-node network with three modules, which is both amenable to numerical analysis and large enough to model cases of interest. 
The network structure is shown in Fig.~\ref{fig:oracle}~(a), and the weights are randomized
to avoid special symmetry. 
As shown in Fig.~\ref{fig:oracle}~(b), some FI eigenvalues of the original inference problem are as small as $ 10^{- 10} $. 
We take $ 5 \times 10^6 $ samples from the distribution each time, so there is no possibility to achieve accurate inference on the eigenvectors with such small eigenvalues. 

We perform numerical optimization of $ \bmh $ with the true $ \mathcal{I} $ and $ \bmj $ in Eq.~\ref{eq:hn}, and provide the resulting optimal field to the learning procedure.
After applying the field, the eigenvalues of FI are increased by orders of magnitude. 
With only two perturbations, the smallest eigenvalue is $ \sim 10^{- 4} $, which is reasonable to infer with our sample size. 
Meanwhile, the previously strong correlations between nodes get 
progressively weakened with more rounds of perturbations, as shown in Fig.~\ref{fig:oracle}~(c). 
We define two measures to quantify the improvement of inference after applying the perturbations. 
The mean estimation error is defined as $ \sum_{i \neq j} |J_{ij} - \tilde{J}_{ij}| / n (n - 1) $.
The edge prediction precision is defined as the ratio of top-$ K $ predicted edges being present in the true $ \bmj $, where $ K $ is the number of edges in $ \bmj $. 
The two measures are shown as a function of the gradient ascent steps in Fig.~\ref{fig:oracle}~(d)(e).

Without perturbation, the average prediction error only slightly decreases with further sampling, and the inferred network has a mix of correct and false links, as shown in Fig.~\ref{fig:oracle}~(f).
Prediction without perturbation produces roughly all positive connections inside each module, and negative connections between modules. 
Intuitively, for strongly coupled networks, we can only know the composition of modules, but not the exact interactions inside and between modules. 
Edge prediction precision under two perturbations ($\sim 1$) was significantly higher than for no perturbation ($0.6$). 
The qualitative structure of the network is successfully
learned with two perturbations. 
Moreover, with five perturbations, we obtain quantitative knowledge of the network. 
The mean estimation error decreases to $ 2 \% $ of the mean interaction strength, and prediction precision rapidly converges to $ 1 $ with a few optimization iterations, 
as shown in Fig.~\ref{fig:oracle}~(c)(e). 

Such improvements are not specific to this network structure and sample size setup.
Numerical experiments show that $ \trinvi $ are reduced by similar folds for random networks, and the fold-change of $ \trinvi $ will not be influenced by network size.
Moreover, the improvements on inference quality are effective for sample sizes as small as a few hundreds, and the effect of perturbation is much more significant than only increasing sample size of the original, unperturbed problem (\textit{SI Appendix}, Text 4, Fig.~S4$-$5). 

\subsection*{Online iterative inference of random networks using inferred fields}

\begin{SCfigure*}[\sidecaptionrelwidth][t]
\centering
\includegraphics[width=11.4cm]{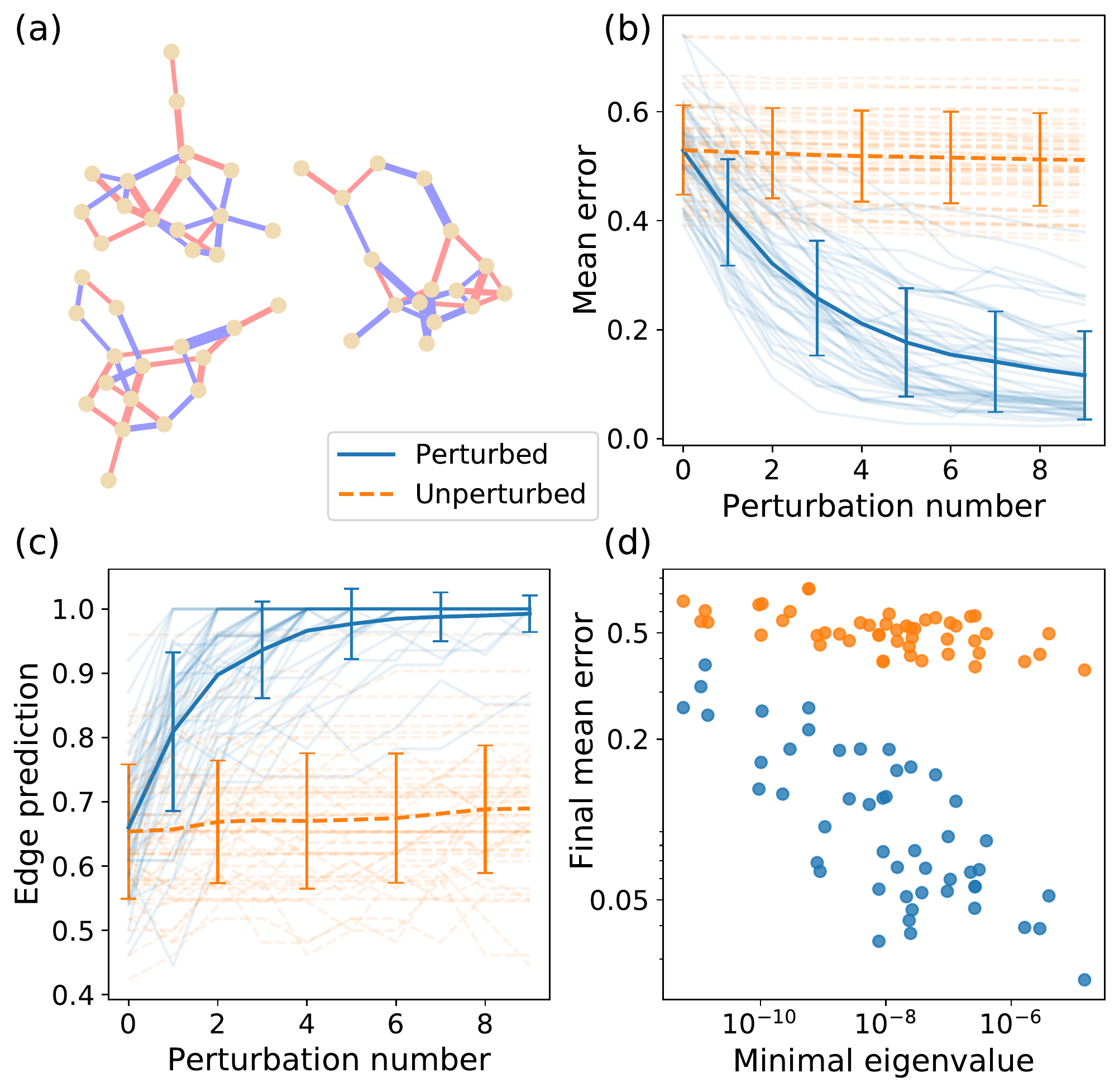}
\caption{\textbf{Inference of random networks using inferred fields.} 
(a) Structure of example 16-node random networks.
(b) Mean estimation error after training is shown as a function of the number of applied perturbations.
The opaque line and error bar represent the mean and standard deviation over the 49 tested random networks. 
The unperturbed group is set as taking the same number of samples but without perturbation. 
The transparent lines in the background show training curves for each individual network.
Legend for (b),(c),(d) shown below panel (a). 
(c) Edge prediction precision 
(defined in Fig~\ref{fig:oracle}~(e)) is shown as a function of the number of applied perturbations. 
Definition of opaque and transparent lines is the same as (b). 
(d) Final mean estimation error as function of the smallest eigenvalue of FI without perturbation. }
\label{fig:random}
\end{SCfigure*}

In previous sections, we showed that perturbations designed by our framework can be applied to general spin-network models, which reduce the sampling complexity by increasing the eigenvalues of FI. 
We now demonstrate good perturbations can be learned online without prior knowledge of the underlying network.
The iteration between model construction and perturbation design is a better representation of most application scenarios. 
In this case, we must infer informative $ \bmh $ using the empirical FI and our current estimate $ \tilde{\bmj} $ in Eq.~\ref{eq:hn}.
We validate the effectiveness of our active learning method in this more challenging context, using 49 randomly generated networks.
Some example network structures are shown in Fig.~\ref{fig:random}~(a).
The smallest eigenvalues of the original inference problem are $\sim 10 ^ {- 7} - 10 ^ {- 10} $, therefore the inference is almost impossible with only $ 5 \times 10^6 $ samples each round. 

The results show that the perturbations discovered using estimation from data can still reveal network structure. 
The inference accuracy of the unperturbed and perturbed systems, are shown in Fig.~\ref{fig:random}~(b)(c) as a function of sampling rounds.
Without perturbation, the mean estimation error almost does not change with more samples, and the edge prediction precision only slightly improves.
In contrast, with inferred perturbations the mean error and edge prediction precision improve significantly with progressively more perturbations. 
For most networks, the edge prediction precision 
eventually converges to $ 1 $.

The quality of inferred perturbation also depends on the eigenvalues of FI. 
The final mean estimation error with 9 rounds of perturbations is strongly correlated with the smallest eigenvalue of FI, as shown in Fig.~\ref{fig:random}~(d).
Meanwhile, for inference without perturbation, the mean estimation error is almost insensitive to the smallest eigenvalue, suggesting that these directions do not provide information on parameter estimation. 
Therefore, even though the current samples is insufficient to precisely infer $ \bmj $, it is sufficient to identify
certain directions where the inference is hard, so that perturbations can be rationally designed to improve accuracy. 

While designing $ \bmh $, we take approximations on Eq.~\ref{eq:hn} with inferred $ \tilde{\bmj} $ and empirical FI. 
We have numerical evidence to demonstrate that the inferred perturbations are still informative under these approximations (\textit{SI Appendix}, Text 5, Fig. S6$-$7). 

\section*{Discussion}

In this paper, we develop an active learning framework to design and analyze perturbation experiments for estimating the structure of complex networks. Our work provides fundamental insight into the physics of active learning. We show that perturbations can break strong correlations between nodes within a network, boosting the eigenvalues of the FI, and thus reducing the sampling complexity of inference by orders of magnitude.  Perturbations provide significant improvements in both qualitative structure prediction and quantitative interaction-strength estimation. 
Our framework combines statistical inference with active exploration, and thus mimics the scientific discovery process. 


An important implication of our work is that active and passive learning have fundamentally different sampling complexity bounds. 
For example, classic results show that the sampling complexity for passive spin network inference increases exponentially with $\bmj$, while in active learning near-constant scaling with $ \bmj $ is achieved in several examples. 
Our results, thus, suggest that scaling relationships and sampling bounds for active learning differ significantly from the passive case, and formal investigation of these bounds is an important topic for future work. 




We see many opportunities for extending and elaborating our framework. 
First, we design perturbations through finding fields that minimize $ \trinvi $ numerically, except for a few analytically tractable networks.
When applied to large networks, the optimization might be computationally intractable, and it would be more efficient if we could design $ \bmh $ directly from $ \bmj $ and $ \mathcal{I} $ without estimating the FI after each hypothetical perturbation. 
Intuitively we would like the perturbation to break strong correlations between nodes, and improve the probability of states that have not been sampled.
Preliminary results on finding $ \bmh $ based on principal components analysis of the correlation matrix show some utility, but further investigation is required to make this approach practical. 

In this manuscript, we consider an active learning framework with full control over the perturbation, the field $ \bmh $.
Generally in applications, our ability to perturb a system might be more constrained. 
For example, perturbations might be constrained in magnitude, be limited in the number of nonzero components, be imprecise, and some nodes might be essential to the system and thus cannot be perturbed. 
Therefore, it might be practically important to adapt our procedure to solve for perturbations that also satisfy a specific set of constraints. Active learning with constrained perturbations will lead to a new set of optimization problems and new performance bounds. 

Finally, we have considered active learning for a specific class of networks, spin networks. However,our work could be extended to other models of complex networks, such as Bayesian networks, networks with hidden nodes, or systems with stochastic dynamics.
In each case, it will be important to understand the physical basis of active learning and to determine how perturbation can quantitatively change fundamental limits on the efficiency of inference. 

We believe our framework provides an approach for closed-loop inference of network models through iterative rounds of observation, model construction and experimentation.  
Our work could lead to practical strategies for inferring the structure of complex networks in physics and biology where large scale perturbations can be applied iteratively. Further, active learning strategies motivate new theoretical directions in machine learning focused on developing formal theories for determining how experimentation can impact the computational and observational efficiency of inferring scientific models.

\matmethods{
A MATLAB~\citep{matlab} implementation of the iterative active learning procedure as well as
all the numerical experiments is made available at \url{https://github.com/JialongJiang/Active_learning_spin}.
Data and Python code to produce all the figures are also included in the repository. 
The experiments are implemented with the following details.  

\subsection*{Learning of network parameters} 
For each round of learning, $ 5 \times 10^6 $ examples are taken from the network by MCMC sampling. 
The optimization is performed by gradient ascent, and $ 5 \times 10 ^3 $ samples are used in each step to estimate the gradient. 
The step size is chosen as $ \eta = \lambda t^{- \alpha} $, where $ \lambda = 0.1 $, $ \alpha \in [0.2, 0.5] $ depending on learning stages. 
To avoid over-fitting, $ \ell_2 $ regularization is used during the training. 

\subsection*{Generation of random networks}

The random networks are generated by cutting off Gaussian random variables. 
Each edge is assigned a weight from the standard normal distribution, and we only keep the weights larger than $ 1.4 $ in magnitude. 
Then all remaining weights are rescaled to make their mean absolute value equal to $ 2.5 $.

\subsection*{Optimization of applied fields}

For $ 16 $-node networks,  true Fisher information was used in the computation. 
When computing the trace of the inverse, an identity matrix with $ 10^{-6} $ weight was added to avoid numerical instability. 
The optimization was performed by the optimization toolbox in MATLAB~\citep{matlab}. 
}

\showmatmethods{} 

\acknow{The authors would like to thank Venkat Chandrasekaran and Andrew Stuart for influential discussions, and Yifan Chen for helpful suggestions. The authors would like to acknowledge support from the Heritage Medical Research Institute (MT), the NIH (DP5 OD012194) (MT), the Natural Sciences and Engineering Research Council (NSERC) Discovery Grant (DAS), and a Tier-II Canada Research Chair (DAS). }

\showacknow{} 

\bibliography{pnas-sample}

\clearpage


\section*{Supplementary material}

**Supporting Information Text**

## 1. Supplementary information on the theoretical framework and inference procedure

**A. General formulation of the framework.** In this section, we develop a general formulation of our framework that can be applied to probabilistic models from the exponential family of distributions with linear parameters. The model class includes systems described by Hamiltonian energy functions.

For an exponential family of distributions on a discrete random variable $\boldsymbol{x}$ with parameter $\boldsymbol{\theta} \in \mathbb{R}^k$ to infer and $\boldsymbol{\eta} \in \mathbb{R}^l$ to control,

$$P(\boldsymbol{x}|\boldsymbol{\theta},\boldsymbol{\eta}) = \frac{1}{Z} \exp\left(-\sum_{i=1}^{k} \theta_i \Psi_i(\boldsymbol{x}) - \sum_{i=1}^{l} \eta_i \psi_i(\boldsymbol{x})\right), \quad [1]$$

where $\Psi_i(\boldsymbol{x}), \psi_i(\boldsymbol{x})$ are functions that specify the distribution. The problem is formulated as facilitating the inference of $\boldsymbol{\theta}$ by controlling $\boldsymbol{\eta}$.

The Fisher information with respect to $\boldsymbol{\theta}$ is

$$\mathcal{I}_{ij} = \left\langle \frac{\partial^2 \log P(\boldsymbol{x}|\boldsymbol{\theta},\boldsymbol{\eta})}{\partial \theta_i \partial \theta_j} \right\rangle = \langle \Psi_i(\boldsymbol{x}) \Psi_j(\boldsymbol{x}) \rangle - \langle \Psi_i(\boldsymbol{x}) \rangle \langle \Psi_j(\boldsymbol{x}) \rangle, \quad [2]$$

which depends only on the control parameter $\boldsymbol{\eta}$. The expectation in both expressions is taken over $P(\boldsymbol{x}|\boldsymbol{\theta},\boldsymbol{\eta})$. Experimental manipulation of $\boldsymbol{\eta}$ modulates the distribution $P(\boldsymbol{x}|\boldsymbol{\theta},\boldsymbol{\eta})$, thus changing the structure of the Fisher information matrix. Intuitively, $\Psi_i(x)$ is the form of interactions in the model, specifically taking the form $s_i s_j$ in the spin-network models. In this general formulation, diagonal entries of FI are variances of $\Psi_i(\boldsymbol{x})$, and off-diagonal entries of FI are covariances between $\Psi_i(\boldsymbol{x})$ and $\Psi_j(\boldsymbol{x})$. The role of perturbations is to increase the diagonal entries, corresponding to exploring more states for each $\Psi_i(\boldsymbol{x})$, while decreasing the off-diagonal entries, corresponding to reducing the covariance between any $\Psi_i(\boldsymbol{x})$ and $\Psi_j(\boldsymbol{x})$.

The numerical procedures developed in the main text can be applied using the expressions Eq. **??** for FI. Specifically, at each round of inference we can estimate the inference parameters $\boldsymbol{\theta}$ using a generalized form of the gradient ascent procedure described in the main text, where the gradient update is now

$$\frac{\partial \log P_n}{\partial \theta_i} = \langle \Psi_i(\boldsymbol{x}) \rangle_n^\circ - \langle \Psi_i(\boldsymbol{x}) \rangle_n + \frac{\partial \log P_{n-1}}{\partial \theta_i}, \quad [3]$$

where, as in the main text, $P_n$ is the posterior distribution given all the samples under the first $n$ fields. $\langle \Psi_i(\boldsymbol{x}) \rangle^\circ$ is the average over the observed samples, and $\langle \Psi_i(\boldsymbol{x}) \rangle$ is the average over the distribution generated by the current parameters. Using the current estimate of $\boldsymbol{\theta}$, we can, then, find new a new perturbation $\boldsymbol{\theta}$ by minimizing the estimated $\operatorname{Tr} \mathcal{I}^{-1}$,

$$\min_{\boldsymbol{\theta}_n} \quad \operatorname{Tr} \mathcal{I}_n^{-1}.$$

In this way, our procedure generalizes to a broader class of networks where interactions might involve higher-order combinations of variables as well as variables that take on a larger set of discrete values.

**B. Derivation of sequential Bayesian estimation.** Here, we derive the Bayesian formula for iterative inference described in the main text Eq. 2. We simply apply the Bayesian formula to the iterative inference process

$$P_n \equiv P(\boldsymbol{J}|\bigcup_{i=1}^{n}\{\boldsymbol{s}\}_i,\boldsymbol{h}_i) = \frac{P(\{\boldsymbol{s}\}_n|\boldsymbol{J},\boldsymbol{h}_n)P(\boldsymbol{J}|\bigcup_{i=1}^{n-1}\{\boldsymbol{s}\}_i,\boldsymbol{h}_i)}{\sum_{\boldsymbol{J}} P(\{\boldsymbol{s}\}_n|\boldsymbol{J},\boldsymbol{h}_n)P(\boldsymbol{J}|\bigcup_{i=1}^{n-1}\{\boldsymbol{s}\}_i,\boldsymbol{h}_i)}, \quad [4]$$

where $P_n$ is the posterior distribution given all samples $\{\boldsymbol{s}\}_i, i = 1, 2, \ldots, n$ taken across the first $n$-rounds of applied fields. Taking the log of $P_n$, we have the decomposition

$$\log P_n = \log \mathcal{L}_n + \log P_{n-1} - \log Z_n, \quad [5]$$

where $\mathcal{L}_n$ is the likelihood function for samples under the $n$-th perturbation, and $Z_n$ is the normalizing constant on the denominator. $\log Z_n$ is a constant, so its gradient vanishes. Thus we obtain

$$\frac{\partial \log P_n}{\partial J_{ij}} = \frac{\partial \log \mathcal{L}_n}{\partial J_{ij}} + \frac{\partial \log P_{n-1}}{\partial J_{ij}} = \langle s_i s_j \rangle_n^\circ - \langle s_i s_j \rangle_n + \frac{\partial \log P_{n-1}}{\partial J_{ij}}. \quad [6]$$

## 2. Analytical solutions for specific networks

**A. Lower bound on trace of inverse of Fisher information.** By the form of FI in the spin network models, the lower bound of $\operatorname{Tr} \mathcal{I}^{-1}$ is

$$\operatorname{Tr} \mathcal{I}^{-1} \geq \frac{p^2}{\operatorname{Tr} \mathcal{I}} \geq p, \quad [7]$$

where the first inequality follows from the harmonic average being smaller than the arithmetic average, and the second inequality by $I_{\{ij\},\{ij\}} = 1 - \langle s_i s_j \rangle^2 \leq 1$. For the quality holds all eigenvalues should equal and the diagonal entries of FI all equal to 1, so the lower bound is achieved if and only if $\mathcal{I}$ is the identity matrix, which means that $\langle s_i s_j s_k s_l \rangle = 0$. The lower bound can be achieved by a uniform distribution on all configurations, but other distributions are also possible. For example, in two-spin networks, there are an infinite number of probability distributions that satisfy the condition.

**B. Optimal Fisher information for two-node inference.** For the two-spin network, the maximum of FI is achieved when $\langle s_1 s_2 \rangle = 0$, which requires

$$\frac{1}{\mathcal{Z}} \left( e^{J+h_1+h_2} + e^{J-h_1-h_2} - e^{-J-h_1+h_2} - e^{-J+h_1-h_2} \right) = 0. \quad [8]$$

Solving the equation gives

$$h_2 = \frac{1}{2} \log \frac{1 - \exp(2J + 2h_1)}{\exp 2J - \exp 2h_1}. \quad [9]$$

Note that the equation is well-defined when $h_1 < J$ or $h_1 > J$, corresponding to the hyperbolic geometry shown in Fig. 1 (b) of main text.

**C. Fisher information of the ferromagnetic Ising chain.** The correlation $\langle s_i s_{i+1} \rangle$ and quadruple correlation $\langle s_i s_{i+1} s_j s_{j+1} \rangle$ can be computed by the transfer matrix method. The partition function without an external field can be written as

$$\mathcal{Z} = \sum_{s_1,\ldots,s_p} \exp\left( J \sum_{i=1}^p s_i s_{i+1} \right) = \operatorname{Tr} P^p, \quad [10]$$

where

$$P = \begin{bmatrix} e^J & e^{-J} \\ e^{-J} & e^J \end{bmatrix}. \quad [11]$$

The eigenvalues of the transfer matrix are

$$\lambda_1 = e^J + e^{-J}, \quad \lambda_2 = e^J - e^{-J}. \quad [12]$$

By symmetry of the Ising chain with periodic boundary conditions,

$$\langle s_i s_{i+1} \rangle = \langle s_1 s_2 \rangle = \frac{1}{\mathcal{Z}} \sum_{s_1,\ldots,s_p} s_1 s_2 \exp\left( J \sum_{i=1}^p s_i s_{i+1} \right) \quad [13]$$

$$= \frac{1}{\mathcal{Z}} \operatorname{Tr}\left( \frac{\partial P}{\partial J} P^{p-1} \right) = \frac{\operatorname{Tr} Q P^{p-1}}{\operatorname{Tr} P^p} = \frac{\lambda_1 \lambda_2^{p-1} + \lambda_2 \lambda_1^{p-1}}{\lambda_1^p + \lambda_2^p},$$

where $Q$ is defined as

$$Q = \frac{\partial P}{\partial J} = \begin{bmatrix} e^J & -e^{-J} \\ -e^{-J} & e^J \end{bmatrix} \quad [14]$$

For the quadruple correlation function, noticing that

$$PQ = QP = \lambda_1 \lambda_2 \operatorname{diag}(1,1), \quad [15]$$

we have

$$\langle s_i s_{i+1} s_j s_{j+1} \rangle = \frac{1}{\mathcal{Z}} \operatorname{Tr}\left[ P^{i-1} Q P^{j-i-1} Q P^{p-j} \right] = \frac{\lambda_1^2 \lambda_2^{p-2} + \lambda_2^2 \lambda_1^{p-2}}{\lambda_1^p + \lambda_2^p}, \quad i \neq j. \quad [16]$$

Then the series expansion at $e^J \to \infty$, then for large $J$

$$\mathcal{I}_{\{i,i+1\}\{j,j+1\}} = \langle s_i s_{i+1} s_j s_{j+1} \rangle - \langle s_i s_{i+1} \rangle \langle s_j s_{j+1} \rangle \quad [17]$$

$$\approx \begin{cases} 4(p-1)\exp(-4J) & i = j \\ 4\exp(-4J) & i \neq j \end{cases}. \quad [18]$$

## 3. Analysis of procedure for three-node networks

To help understand the mathematical framework and its implications, we provide a global analysis of three-node networks and a detailed analysis of optimal perturbations for two specific three-node networks. In general, there are 7 different topologies of connected three-node networks (up to a permutation of node labels), as shown in Fig. S1 (a), where the topology is defined as the signed edge connectivity. For three-node networks, the optimal perturbation can be found numerically by exhaustive grid search of all possible field directions and magnitudes, and so the networks provide a tractable set of examples in which we can comprehensively explore the impact of perturbation on inference.

Setting the absolute value of all interactions equal to 2, $\operatorname{Tr} \mathcal{I}^{-1}$ for each topology without perturbation and with one numerically optimal perturbation is shown in Fig. S1 (b). From this analysis, we can draw two general conclusions. First, the difficulty of inference, as represented by $\operatorname{Tr} \mathcal{I}^{-1}$, depends on network topology. Networks 2, 4 and 6 are not fully connected and have smaller $\operatorname{Tr} \mathcal{I}^{-1}$ compared to other networks without perturbation. Therefore, these networks are intrinsically "easier" to learn by observation. Second, network topology also impacts the optimal $\operatorname{Tr} \mathcal{I}^{-1}$ with perturbation. All of the incompletely connected networks, as well as networks 3 and 7, achieve the lower bound $\operatorname{Tr} \mathcal{I}^{-1} = 3$ after a single perturbation. Conversely, the perturbation only decreases $\operatorname{Tr} \mathcal{I}^{-1}$ of networks 1 and 5 from $10^3$ to $10^2$. Therefore networks 3 and 7 are "easy" to infer

with optimal perturbation, while networks 1 and 5 are "hard" even with one optimal perturbation, which demonstrates the necessity of performing multiple rounds of perturbations for certain classes of networks. Thus, network topology determines the behavior of $\text{Tr}\,\mathcal{I}^{-1}$ by defining the underlying energy landscape and thus form of the distribution on configurations. We take networks 3 and 1 for detailed analysis.

By the symmetry of the system, FI is the same if the sign of the field is flipped, so we can set $h_1 > 0$ without loss of generality. Then the applied field can be parametrized as

$$\bm{h} = |\bm{h}| \left[ \sqrt{1 - h_2^2 - h_3^2}, h_2, h_3 \right]^T , \qquad [19]$$

where $|\bm{h}|$ is the Euclidean norm of $\bm{h}$. Given the direction of perturbation, the minimum of $\text{Tr}\,\mathcal{I}^{-1}$ over $|\bm{h}|$ can be visualized as a heatmap of $[h_2, h_3]$, as shown in Fig. S2 (b) and Fig. S3 (b).

For the network in Fig. S2 (a), the optimal FI achieves the lower bound $\text{Tr}\,\mathcal{I}^{-1} = 3$, and the optimal perturbation is approximately $[2J, -2J, -2J]$. However, for the network in Fig. S3 (a), the minimum of $\text{Tr}\,\mathcal{I}^{-1}$ is far larger than 3. As shown in Fig. S2 (b) and Fig. S3 (b), the direction of perturbation is crucial to the resulting optimal FI. The eigenvectors and eigenvalues without and with perturbation are shown in Fig. S2 (c) and Fig. S3 (c). These eigenvectors and eigenvalues can be interpreted with reference to the network structure. For example, the smallest eigenvalue in inferring network 3 has eigenvector corresponding to the signed edge connectivity. This is because increasing the edge intensity proportionally does not change the distribution much, and therefore edge intensity is hard to determine. The effect of the perturbation on the distribution can be visualized by comparing the energy of all configurations, as shown in Fig. S2 (e) and Fig. S3 (e). Blue (orange) dots represent energies without (with) perturbation. The effect of perturbation is to create multiple low-energy states, in other words, to make some "informative" configurations have high probabilities.

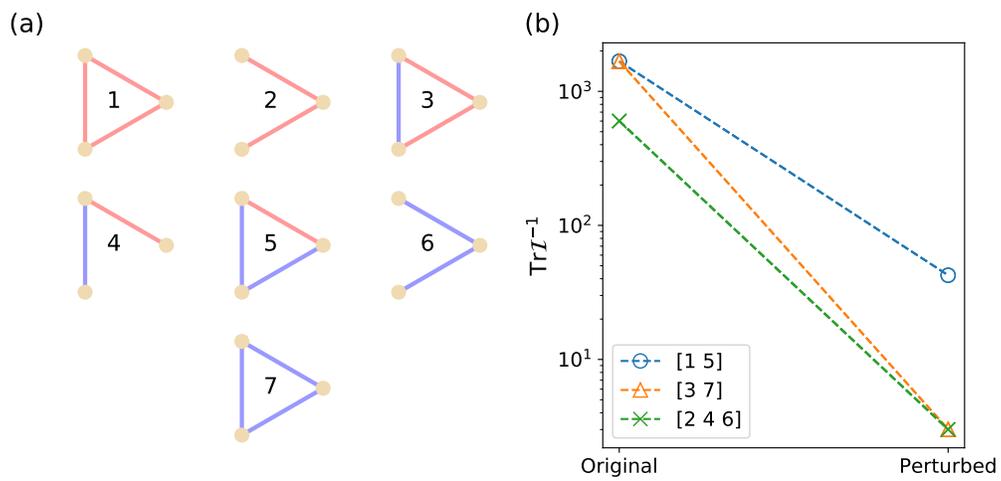

**Fig. S1. The effect of perturbation on three-node networks.** (a) All possible topologies of connected three-node networks. Red (blue) edges represent positive (negative) interactions. (b) All interaction strength are set to $2$. Based on $\mathrm{Tr}\,\mathcal{I}^{-1}$ of the original problem and optimally perturbed problem, these networks can be classified into three groups. Legend shows the indexes of networks in each group.

Jialong Jiang, David A. Sivak and Matt Thomson

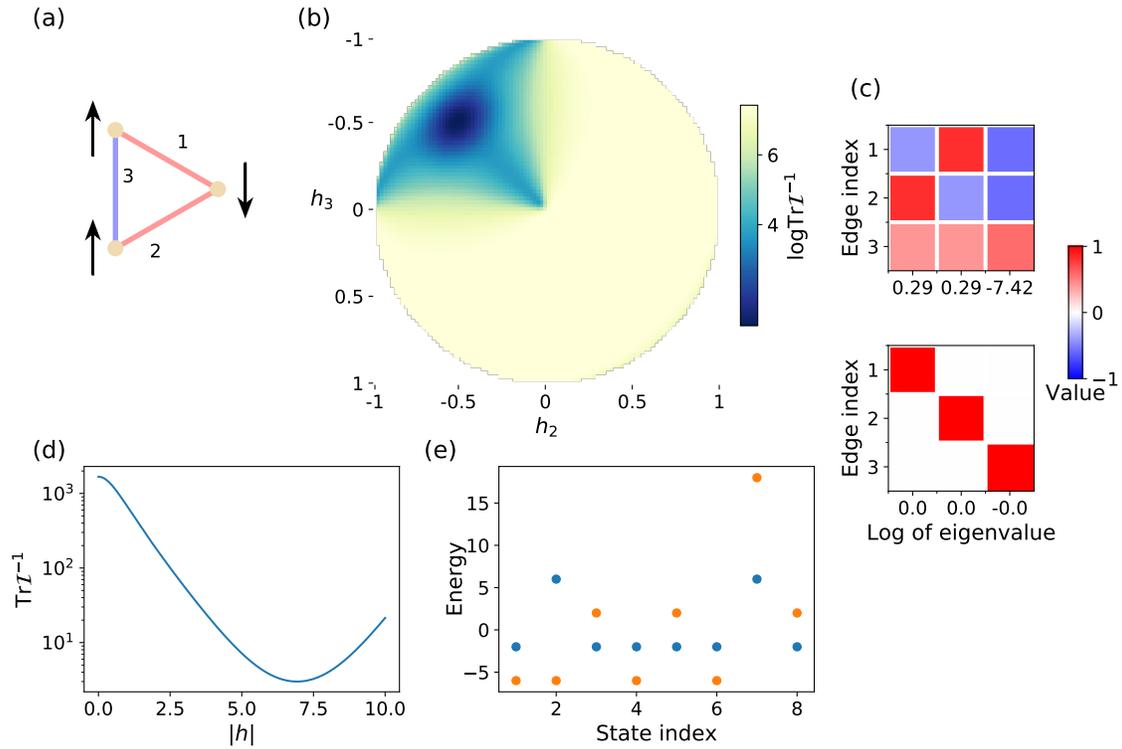

**Fig. S2. Optimal perturbation of an "easy" three-node network.** (a) The structure of the network and the optimal perturbation. Interaction strengths are set to $2$, and the sign is indicated by the color as in Fig. S1. Applied fields are illustrated by the direction and length of arrows. (b) Minimum of $\text{Tr}\,\mathcal{I}^{-1}$ on the perturbation direction specified by $h_2$ and $h_3$ as in Eq.19. (c) Eigenvectors and eigenvalues of FI without and with the optimal perturbation. Each eigenvector is represented as a column in the heatmap, and the logarithm of the corresponding eigenvalue is shown below. (d) $\text{Tr}\,\mathcal{I}^{-1}$ as a function of $|h|$ along the optimal perturbation direction. (e) The energy of each state without and with optimal perturbation. Blue (orange) dots are energies without (with) perturbation.

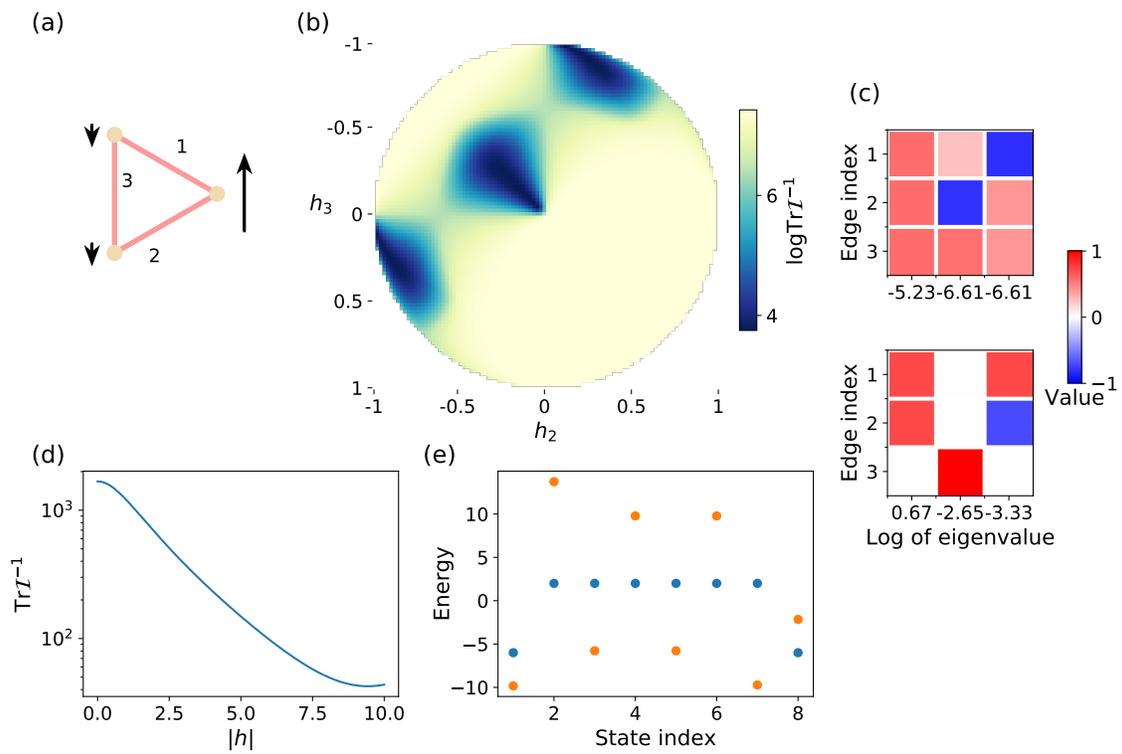

**Fig. S3. Optimal perturbation of a "hard" three-node network.** All other captions are the same as Fig. S2. The optimal perturbation has large magnitude and is hard to illustrate, so $\ell_2$ regularization is used to define the optimal perturbation.

Jialong Jiang, David A. Sivak and Matt Thomson

## 4. Scaling properties of active learning with network size and sample size

In the numerical results section of the main text, we illustrate that suitable perturbations can reduce the sampling complexity of the inference problem by orders of magnitude, thus improving the inference accuracy and efficiency. In this section, we illustrate how the results depend on two key parameters, network size and the number of samples. In particular, the fold-change of $\operatorname{Tr} \mathcal{I}^{-1}$ due to perturbations does not deteriorate with larger network size and harder inference problems. Moreover, the improvements of inference quality are observed with sample sizes as small as 500, and applying perturbation improves inference much more significantly than only increasing sample size of the original problem.

Numerical experiments are performed on random networks that are generated by assigning standard Gaussian interaction strength to each edge and retaining edges with absolute interaction strength larger than a threshold, so that the appearance probability for each edge is 0.25. All interaction strengths are then rescaled to have mean absolute value 2.5. Perturbations are designed with the knowledge of exact $\boldsymbol{J}$.

$\operatorname{Tr} \mathcal{I}^{-1}$ of such random networks with 8, 12, and 16 nodes under different number of perturbations are shown in Fig. S4 (a). $\operatorname{Tr} \mathcal{I}^{-1}$ without perturbation scales exponentially with the network size, and its variance is large across random networks. Perturbations reduce both mean and variance of the distribution of $\operatorname{Tr} \mathcal{I}^{-1}$, as seen in Fig. S4 (b) and Fig. S5 (a)(d)(g). The fold changes of $\operatorname{Tr} \mathcal{I}^{-1}$ are shown in Fig. S4 (c). The fold changes of $\operatorname{Tr} \mathcal{I}^{-1}$ are more significant for larger networks, even though the original inference problem is harder.

The mean interaction strength prediction error and structural edge prediction accuracy, as a function of sample size, are shown in Fig. S5. For very small sample sizes (5, 50), the perturbed problem has higher edge prediction accuracy, but also higher mean estimation error, while for sample size larger than 500 both inference quality measures are improved by perturbation. Increasing sample size can improve inference quality, but the effect is weak at large-enough sample size and overall, perturbations are more effective than increasing sample size. 5 perturbations with sample size 500 can give better inference than the unperturbed problem with $5 \times 10^6$ samples.

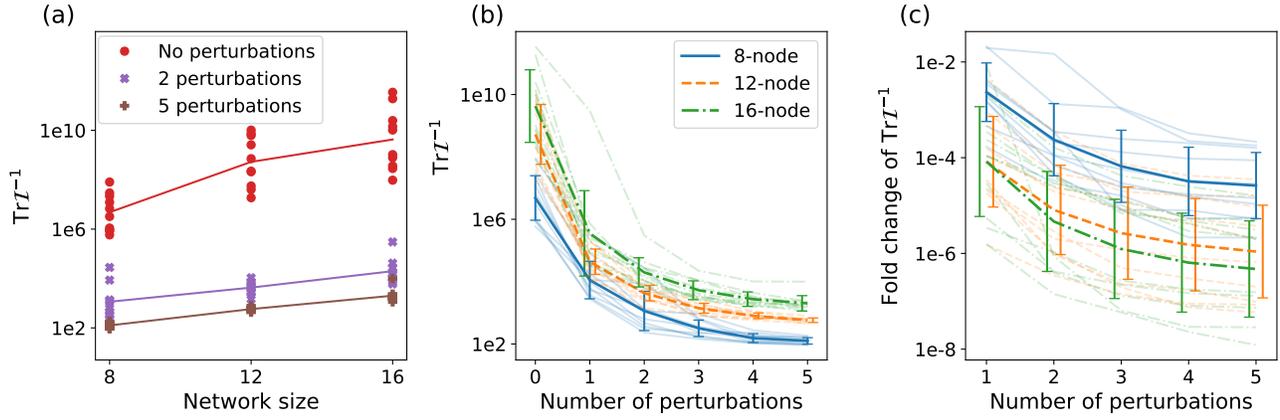

**Fig. S4. Trace of inverse of FI, as a function of network size and number of perturbations.** (a) Scaling of $\operatorname{Tr} \mathcal{I}^{-1}$ with network size, under different numbers of perturbations. Lines connect the average of 10 random networks for each size in logarithm scale. (b) $\operatorname{Tr} \mathcal{I}^{-1}$ decreases with number of perturbations. The opaque line and error bar represent the mean and standard deviation, calculated in the logarithm scale. The transparent background lines show inference performance for all individual networks. (c) Fold change of $\operatorname{Tr} \mathcal{I}^{-1}$ relative to that without perturbation, as a function of number of perturbations. The legend and meaning of lines are the same as (b).

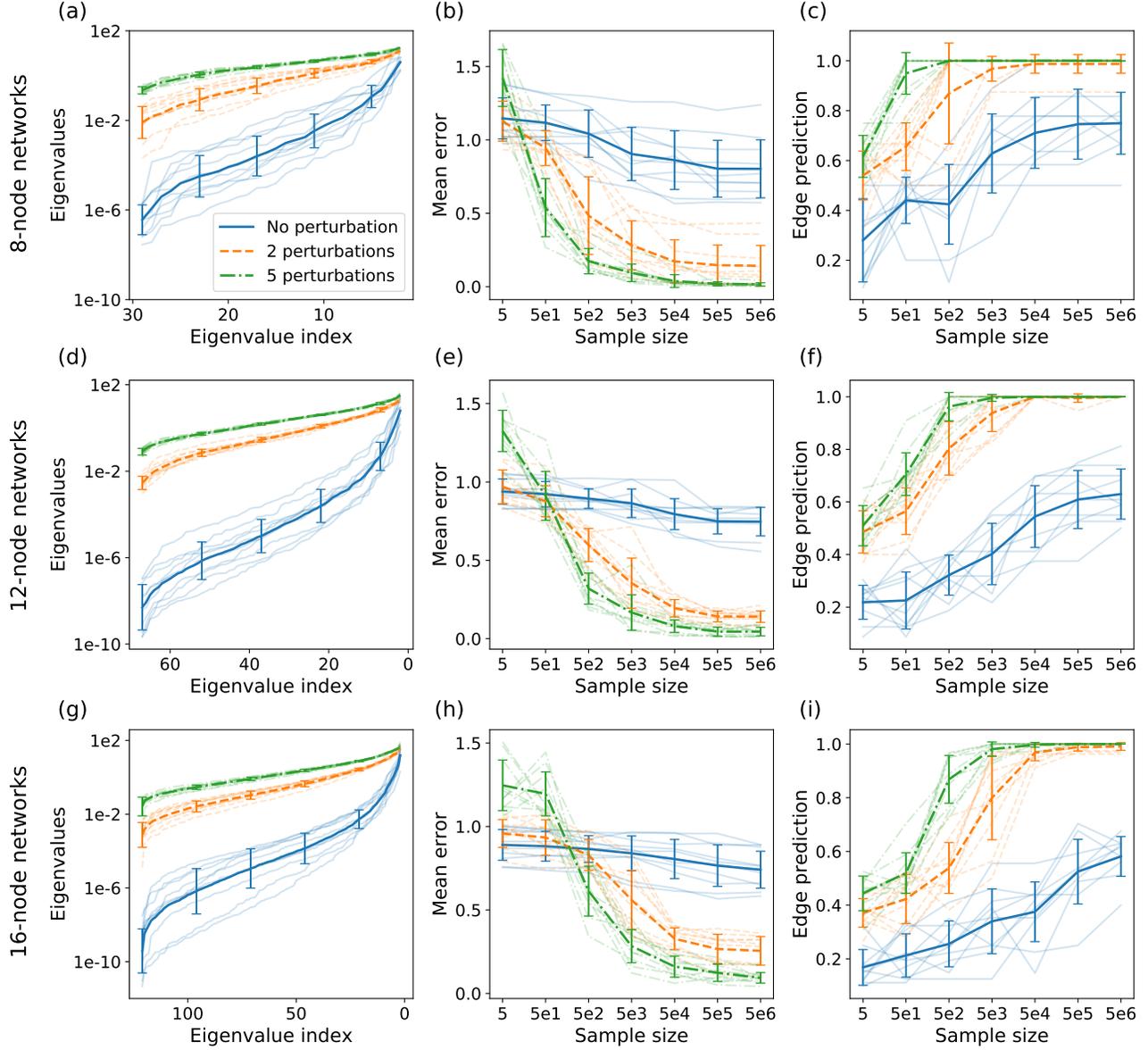

**Fig. S5. Scaling of random-network FI eigenvalues and inference quality with sample size.** Top row (a)(b)(c): 8-node random networks. Middle row (d)(e)(f): 12-node random networks. Bottom row (g)(h)(i): 16-node random networks. Left column (a)(d)(g): Eigenvalues of FI with given number of perturbations. Eigenvalues of 10 random networks are shown in transparent line, and opaque line and error bar are mean and standard deviation in logarithm scale. The legend is shared for all panels. Middle column (b)(e)(h): Final mean estimation error of $J_{ij}$ as a function of sample size. Data for each random networks are shown in transparent line, and opaque line and error bar are mean and standard deviation. in (b)(c). Right column (c)(f)(i) Edge prediction precision as a function of sample size. Line interpretation is the same as Middle column.

Jialong Jiang, David A. Sivak and Matt Thomson

## 5. Effect of using inferred interaction matrix and empirical Fisher information

In the main text section of designing perturbation online, several approximations are made with some implicit assumptions. We argue that these approximations are valid in the sense of finding good perturbations.

**A. Convergence of empirical Fisher information spectrum.** The first approximation is that empirical FI is used instead of the true FI. Results in random matrix theory have shown that the empirical FI converges to the true FI with increasing number of samples, and that the convergence rate for different eigenvectors is proportional to the exponential of the eigenvalues. We numerically verify that empirical FI well approximates the true FI in the direction of eigenvectors with large eigenvalues, as shown in Fig. S6.

As shown in Fig. S6 (a)(c)(e), the estimated FI gives a good estimate of the eigenvalues of the true FI when the eigenvalue is larger than $\sim 10^{-5}$. This is related to our sample size $5 \times 10^6$. The samples do not contain sufficient information about very small eigenvalues, so the corresponding eigenvalues in the estimated FI are close to the magnitude of numerical error.

The estimation quality of eigenvectors can be quantified by computing the inner product between any eigenvectors in the true FI and estimated FI. The squared inner product is shown in Fig. S6 (b)(d)(f). The heatmap is close to diagonal at the bottom-right corner, showing that the estimation of eigenvectors is relatively precise for eigenvectors with large eigenvalues. As the number of perturbations increases, the heatmap becomes more concentrated on the diagonal, which means the estimation of eigenvectors becomes more accurate.

**B. Comparison of designed perturbation with inferred interaction matrix.** The other approximation is that we compute $\mathcal{I}_n$ using the estimate $\tilde{J}$ in place of the true $J$. We show that inferred $J$ is still informative for designing perturbations. We apply perturbations on a 16-node random networks that are either (1) random Gaussian variables with standard deviation equal to the average interaction strength, (2) designed using inferred $\tilde{J}$ and empirical FI, or (3) designed using true $J$ and true FI. The results are shown in Fig. S7.

The inferred $\tilde{J}$ is obtained by performing inference on the unperturbed network. The comparisons between true $J$ and inferred $\tilde{J}$ are shown in Fig. S7 (a). Inferred $\tilde{J}$ preserves the major structures in $J$ but also contains many false interactions. We design the perturbation using the three different methods, and add the obtained new FI to the FI of the original inference problem, $\operatorname{Tr} \mathcal{I}^{-1}$ of which is shown in Fig. S7 (b). Even though the perturbations designed with inferred $\tilde{J}$ are not as good as those with accurate $J$, they are still better than the random perturbations. The designed perturbations are visualized in Fig. S7 (b), showing some qualitative agreement of features between them.

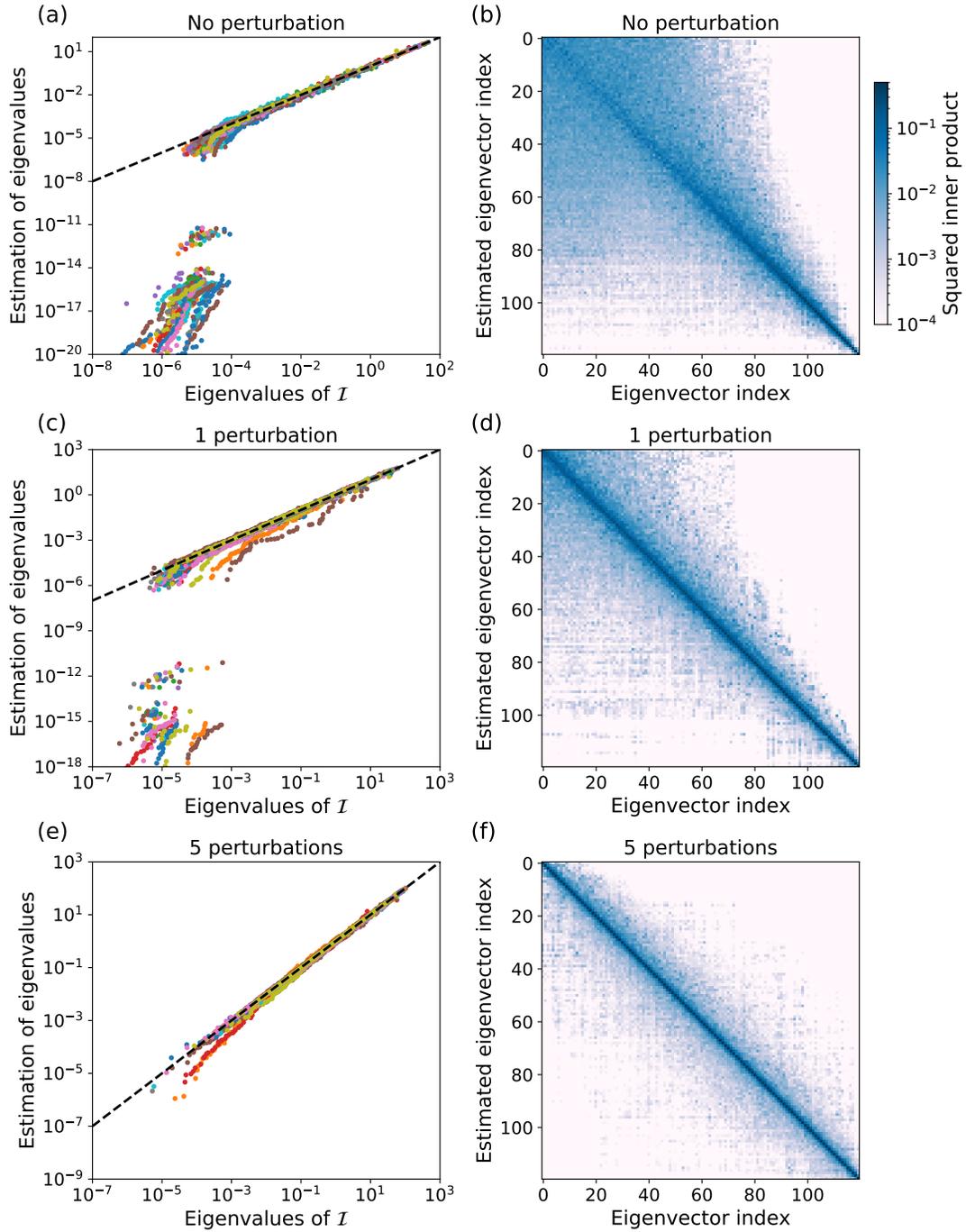

**Fig. S6. Spectrum of estimated FI.** (a) The relation between ranked eigenvalues of true FI and estimated FI is shown as a scatter plot. Each color is a different network among the 49 tested random networks. The black dashed line is a reference $x = y$ line. (b) The squared inner product between eigenvectors of estimated FI and exact FI is shown as a heatmap. The corresponding eigenvalues of eigenvectors increase from left to right, and from top to bottom. (c)(d) The same plot as (a)(b) for the estimated FI and true FI with one round of perturbation. (e)(f) The same plot as (a)(b) for the estimated FI and true FI with 5 rounds of perturbation.

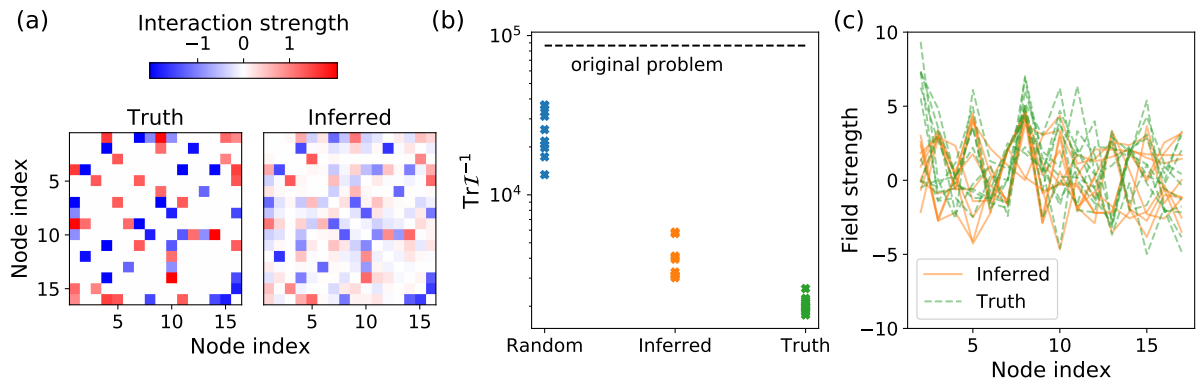

**Fig. S7. Effect of using inaccurate $J$ and empirical FI.** (a) True $J$ and inferred $J$ are shown as heatmaps. (b) Comparison of $\mathrm{Tr}\,\mathcal{I}^{-1}$ after perturbation by random Gaussian fields, by optimizing with inferred $\tilde{J}$, or by optimizing with true $J$. Each perturbation type has 10 repetitions. $\mathrm{Tr}\,\mathcal{I}^{-1}$ of the original problem is shown as the dashed line. (c) The perturbations generated using inferred $\tilde{J}$ and true $J$.



\end{document}